\documentclass[prl,showpacs,twocolumn,amsmath,amssymb]{revtex4}
\usepackage{graphicx}
\usepackage{color}

\def\beq{\begin{equation}}
\def\ee{\end{equation}}
\def\bi{\begin {itemize}}
\def\ei{\end{itemize}}

\def\lsim
{\protect \raisebox{-0.75ex}[-1.5ex]{$\;\stackrel{<}{\sim}\;$}}
\def\gsim
{\protect \raisebox{-0.75ex}[-1.5ex]{$\;\stackrel{>}{\sim}\;$}}
\def\lsimeq
{\protect \raisebox{-0.75ex}[-1.5ex]{$\;\stackrel{<}{\simeq}\;$}}
\def\gsimeq
{\protect \raisebox{-0.75ex}[-1.5ex]{$\;\stackrel{>}{\simeq}\;$}}

\def\njp{{n_j}^+}
\def\njm{{n_j}^-}

\def\beq{\begin{equation}}
\def\ee{\end{equation}}

\def\bi{\begin {itemize}}
\def\ei{\end{itemize}}

\def\weqmn{w^{\rm eq}_{mn}}

\def\wmn{w_{mn}}
\def\wnm{w_{nm}}

\def\dmn{d_{mn}}

\def\dnm{d_{nm}}

\begin{document}

\title{ Generalized Einstein or Green-Kubo relations for active biomolecular
transport}
\author{Udo Seifert
}

\affiliation{
{II.} Institut f\"ur Theoretische Physik, Universit\"at Stuttgart,
  70550 Stuttgart, Germany}
\begin{abstract}

For  driven Markovian dynamics on a network of (biomolecular) states,
the generalized mobilities, i.e., the response of any current to
changes in an external parameter, are expressed by an integral over
an appropriate current-current correlation function and thus related to
the  generalized diffusion constants. As only input, a local detailed
balance condition is required typically even valid for biomolecular
systems  operating deep in the non-equilibrium regime.

\end{abstract}

\maketitle

\def\lsim
{\protect \raisebox{-0.75ex}[-1.5ex]{$\;\stackrel{<}{\sim}\;$}}

\def\gsim
{\protect \raisebox{-0.75ex}[-1.5ex]{$\;\stackrel{>}{\sim}\;$}}

\def\lsimeq
{\protect \raisebox{-0.75ex}[-1.5ex]{$\;\stackrel{<}{\simeq}\;$}}

\def\gsimeq
{\protect \raisebox{-0.75ex}[-1.5ex]{$\;\stackrel{>}{\simeq}\;$}}

{\sl Introduction.--} Close to equilibrium, transport coefficients
like the mobility, conductivity or viscosity, quantifying the response
of a system to an external field or perturbation, can be expressed by
equilibrium correlation functions. The Stokes-Einstein
relation between the mobility and the diffusion constant of a
spherical particle is arguably the oldest and best known example of
such a Green-Kubo relation \cite{kubo}. Both mobility and diffusion constant
are still well-defined even for a non-equilibrium steady state (NESS)
of an open or driven system in which stationary currents lead to 
permanent dissipation. In such a state, the Stokes-Einstein relation
no longer holds true. The difference between diffusion constant
and mobility, however, can be 
expressed by an integral over an  experimentally
 measurable  correlation function \cite{blic07}.

In the present paper, we investigate the relation between a
mobility or transport coefficient and the corresponding dispersion
or fluctuations for any current in an arbitrary driven system with 
the special focus
on biomolecular transport like the one mediated by molecular motors
or ion channels and pumps. The essential characteristics of such transport is that
even though the system is driven, typically by non-balanced chemical
reactions involving ATP, it takes place in a well-defined thermal 
environment. This fact imposes a constraint on the ratio between
forward and backward rates for any mesoscopic transition that will
allow us 
to express the difference between
 mobility and dispersion in a physically transparent way.
On the technical level, we build on the recent derivation of
a general fluctuation-dissipation theorem for NESSs 
\cite{marc08,baie09,pros09,seif09}. 
By directly working in the NESS, our approach is
complementary 
to work that invokes the fluctuation theorem for deriving
non-linear response coefficients in higher order expansions around
equilibrium \cite{andr07b,astu08}. 
Moreover, it goes  beyond similar relations 
obtained for genuine 
diffusive spatial transport \cite{spec06,chet09}
since we require no 
Euclidean metric and hence the notion of a locally co-moving frame
is not available. Our results will therefore be applicable 
not only to any discrete model for a molecular motor or ion pump
(see, e.g., 
\cite{lau07a,kolo07,liep07a,kim09} and references therein) but also to 
driven (bio)chemical reaction networks and their response to changing 
chemical conditions \cite{heue06,schm06}. As a simple illustration will 
show, a misguided rewriting of our additive relationship between mobility 
and dispersion in terms of a multiplicative ``effective temperature''
could easily lead even to negative values for the latter as found for 
various active biomolecular systems, see, e.g.,  \cite{mart01,kiku09}.

{ \sl System.--}
We describe the system by a set of discrete states $\{n\}$.
At time $t$, the system is in a state $n(t)$ jumping at  discrete 
times $t_j$ from state $\njm$ to 
state $\njp$. A transition between state $m$ and state $n$ occurs 
with a rate $ \wmn$.
With each transition $m \to n$, we associate  transport
of a quantity $d^\alpha_{mn}= - d^\alpha_{nm}$ leading to a microscopic current
\beq
 j_\alpha(t) \equiv 
  \sum_j \delta (t-t_j) d_{\njm\njp}^\alpha .
\label{eq:j}
\ee 
The transition rates between the states depend on a set of
 external parameters
$\{h_\beta\}$.
We make no particular assumptions on the parameter dependence of the
individual transition rates but only require that the ratio between forward and
backward rates obeys the typical ``local detailed balance'' (LDB) condition
\beq
\frac{\wmn(\{h_\beta)\}}{\wnm(\{h_\beta\})} = \frac{\wmn(\{0\})}{\wnm(\{0\})}
\exp [\sum_\beta h_\beta d^\beta_{mn}/T] ,
\ee
which implies for the logarithmic derivatives, or ``sensitivities'',
$
r^\beta_{mn} \equiv T \partial_{h_\beta}\ln \wmn
$,
the crucial  relation
\beq r^\beta_{mn}-r^\beta_{nm}= d^\beta_{mn}.
\label{eq:sym}
\ee 
Here, and throughout the paper, we set Boltzmann's constant $k_B\equiv 1$.
Examples for  pairs of an external parameter $h_\alpha$ and a
conjugate distance $d^\alpha$ are 
(i) force $f$ and spatial distance $d$,
(ii) chemical potential $\mu_\alpha$  and number $d^\alpha$
of consumed (or, if negative, produced)
molecules of type $\alpha$ (like ATP and ADP)  and (iii) potential
difference $\Delta \phi$   and transported 
electrical charge $q$. These choices are  relevant
 to  molecular motors (i-ii) and ion pumps (ii-iii),
respectively. In all these cases, the LDB condition is usually assumed
not only for small deviations from equilibrium but also for finite
values of the fields $\{h_\beta\}$.

For constant external parameters $\{h_\beta\}$, the system reaches a 
stationary state in which $p_m\equiv \langle \delta_{n(t)m}\rangle $
denotes the probability to find it in the particular 
state $m$. Throughout the paper,
the brackets $\langle 
... \rangle$ denote averages in this stationary state. If the system operates
in a genuine NESS at least one pair of directed probability
currents
\beq
K_{mn}\equiv p_m\wmn -p_n \wnm = -K_{mn}
\ee
is non-zero. Consequently, some of the currents have a
non-zero mean
\beq
j_\alpha \equiv \langle j_\alpha(t)\rangle = \sum_{mn} p_m\wmn d^\alpha_{mn}
= \sum_{mn} K_{mn} d^\alpha_{mn}/2.
\ee

We will need a second type of current derived from a local
variable
\beq
\nu_\alpha(t)  = \sum_m \delta_{n(t)m}\nu_m ^\alpha{~~~\rm with ~~~}
\nu_m^\alpha \equiv   \sum_{k} K_{mk} r_{mk}^\alpha /p_m  
\label{eq:nu}
\ee which could be called a ``sensitivity-weighted''  current.
It generalizes the mean local velocity found in this context
for Langevin systems \cite{spec06} to arbitrary networks.
Positive contributions to $\nu_\alpha(t)$ arise from links for which
the directed probability current and the sensitivity have the
same sign. The dimension of $\nu_\alpha(t)$ justifies to call it a current.
Moreover, its mean is equal to the ordinary current since
$
\langle \nu_\alpha(t)\rangle  =   \sum_{mn} K_{mn} r^\alpha_{mn}  = j_\alpha
$ where we use (\ref{eq:sym}) from above.

{\sl Generalized Green-Kubo relations.--}
The aim of generalizing the Einstein or Green-Kubo relations to
non-equilibrium processes requires that we express both 
 the generalized  diffusion constants, or dispersions, and the 
generalized mobilities 
 by correlation functions 
involving currents. The dispersions given by
\beq
D_{\alpha\beta} \equiv \lim_{t\to \infty} \frac {1}{2t}\int_0^t dt'\int_0^t dt''
\langle (j_\alpha(t')-j_\alpha)(j_\beta(t'')-j_\beta)\rangle  
\ee characterize the integrated fluctuations around the mean currents.
By isolating the diagonal in this
double integral and exploiting stationarity,
we can rewrite the dispersions as
\beq
D_{\alpha\beta} = \int_{0^+}^ \infty dt\langle (j_\alpha(t)-j_\alpha)
(j_\beta(0)- j_\beta)\rangle + 
D^{\rm  loc}_{\alpha\beta} .
\label{eq:Dfirst}
\ee
The lower boundary $0^+$ at the integral indicates that no
delta-like contributions at $t=0$ should be picked up since those
are captured by the time-local contribution
\begin{eqnarray}
D^{\rm loc}_{\alpha\beta} &\equiv& \lim_{\epsilon\to 0}(1/2\epsilon) \int_{-\epsilon/2}^{\epsilon/2}
dt \langle j_\alpha(t) j_\beta(0)\rangle\\ &=&
(1/2) \sum_{mn}p_m\wmn d^\alpha_{mn}
d^\beta_{mn}  .
\label{eq:Dloc}
\end{eqnarray}

The generalized mobilities 
$\kappa_{\alpha\beta} \equiv  \partial_{h_\beta} j_\alpha$ quantify 
the dependence of the mean current on an external parameters. 
As our main result, we will prove below that they can also
be expressed by an integral involving a correlation function
of the currents just introduced
and a local term in the form
\beq
\kappa_{\alpha\beta} =  \int_{0^+}^ \infty dt \langle j_\alpha(t) 
(j_\beta(0)-\nu_\beta (0))\rangle/T + 
\kappa_{\alpha\beta}^{\rm loc}  
\label{eq:kappamain}
\ee
where
\beq
 \kappa^{\rm loc}_{\alpha\beta} 
 \equiv  \sum_{mn} d^\alpha_{mn} p_m\wmn r^\beta_{mn}/T .
\label{eq:kloc}
\ee 
Hence, the difference between the dispersion and  mobility tensors
can be expressed  as
\beq
I_{\alpha\beta} = D_{\alpha\beta}  - T \kappa_{\alpha\beta} = 
\int_0^\infty dt \langle j_\alpha (t)( \nu_\beta(0)-  j_\beta)\rangle
+ I^{\rm loc}_{\alpha\beta}
\label{eq:I}
\ee
with the local contribution 
\beq
I^{\rm loc}_{\alpha\beta} = D^{\rm loc}_{\alpha\beta} -T \kappa^{\rm
  loc}_{\alpha\beta}= 
 - \sum_{m<n}d^\alpha_{mn}K_{mn}( r^\beta_{mn}+r^\beta_{nm})/2  ,
\label{eq:Ilocab}
\ee
where the notation $\sum_{m<n}$ indicates that each
link is counted only ones.

In equilibrium, $\nu_\beta(0), j_\beta$ and $K_{mn}$ all vanish identically,
and hence $I_{\alpha\beta} = 0 $. Our representation
makes the ``violation'' of the Einstein or Green-Kubo in a NESS apparent and 
provides a  physically transparent expression for the difference between
dispersions and mobilities.

{\sl Molecular motor.--} As an illustration of the general framework we
consider any discrete state model of a molecular motor. 
A transition from state $m$ to state $n$ 
 may either advance the motor 
a spatial distance $\dmn=-\dnm$, or be associated with a
chemical reaction of the type
$
\sum_\alpha r^\alpha_{mn} A_\alpha \to \sum_\alpha s^{\alpha}_{mn}
A_\alpha ,
$ or contain both. 
 The index $\alpha = t,d,p$ labels the chemical species
ATP, ADP and P$_{\rm i}$, respectively,  and $r^\alpha_{mn}$ and
$s^\alpha_{mn}
(=r^\alpha_{nm})$ are the corresponding
stochiometric factors for the forward and backward reaction. 
For
each transition and each species  a 
``chemical distance'' 
\beq
d^\alpha_{mn}\equiv r^\alpha_{mn}-s^\alpha_{mn} =  r^\alpha_{mn}-
r^\alpha_{nm}= -d^\alpha_{nm} ,
\label{eq:dchem}
\ee
 denotes the
number of consumed (or, if negative, produced) molecules of type $\alpha$.
The chemical species are provided at externally  controlled
  concentrations $c_\alpha$.
For any motor and  no applied external force ($f=0$) there are    
 concentrations $c_\alpha^{\rm eq}$
at which the motor is in equilibrium with its thermal
and chemical environment. Assuming ideal behaviour,
 the 
concentrations are
linked to the chemical potentials by
$
\mu_\alpha = \mu_\alpha^{\rm eq} + T \ln c_\alpha/c^{\rm eq}_\alpha .
$
If, still at $f=0$, the chemical potentials deviate from their equilibrium
value, the transition rates are modified according to the usual mass
action law kinetics,
\beq
\wmn =\weqmn \exp \sum_\alpha \Delta\mu_\alpha r^\alpha_{mn}/T,
\ee
where $\Delta \mu_\alpha\equiv  \mu_\alpha -  \mu_\alpha^{\rm eq}$.
Note that the dependence of these rates on the chemical potentials
($h_\alpha=\Delta \mu_\alpha$)
obeys (\ref{eq:sym}) which justifies a posteriori
to denote the stochiometric coefficients by $r^\alpha_{mn}$.
We make no particular assumptions on the force dependence of the
individual transition rates but require that the ratio between forward and
backward rates obeys, as usually assumed, the LDB condition
\beq
\frac{\wmn(f)}{\wnm(f)} = \frac{\wmn(0)}{\wnm(0)}\exp (f\dmn /T) .
\ee
Hence,  the sensitivities $
r_{mn} \equiv T \partial_f\ln \wmn
$ obey
the relation (\ref {eq:sym}).

\def\Te{T^{\rm eff}}
\def\tp{\theta_+}
\def\tm{\theta_-}

For a simple but still instructive specific example, we consider a
``one-state'' ratchet model where the forward rate (driven by ATP hydrolysis)
and the backward rate (synthesizing ATP from ADP and P$_{\rm i}$) 
are given by
\beq
w_+ = w_+^{\rm eq} \exp [(\Delta \mu^t + f\tp d)/T]
\ee
and
\beq
w_-=w_-^{\rm eq} \exp [(\Delta \mu^d + \Delta \mu^p - f\tm d)/T] ,
\ee
respectively.
 The load sharing factors $\tp$ and $\tm$ with $\tp+\tm =1 $ guaranteeing
the LDB condition (\ref{eq:sym}) are
related to the distance of the activation barrier  in forward and
backward direction, respectively \cite{kolo07}.  

Since in this model all sites
are physically equivalent but only spatially translated a distance $d$, 
there are no 
current correlations, so that only the local terms contribute. With
$j=d(w_+-w_-)$, 
the ordinary spatial mobility becomes 
$
\mu \equiv \partial_fj = d^2 [ \tp w_+
+  \tm w_-]/T, 
$
and   the corresponding diffusion coefficient 
$
D= (1/2) d^2 [w_+ + w_-]  .
$
The difference $
I = -d^2 (w_+-w_-)(\tp -\tm)/2 
$ vanishes not only in equilibrium ($w_+ = w_-$)
but even in a NESS for a symmetric barrier ($\tp=\tm=1/2$). 

Expressed in terms of an effective temperature, 
\beq
\Te\equiv D/\mu = T+ I/\mu=\frac{T(\rho+1)}{2 (\tp  \rho + 1-\tp)}  ,
\ee
where $\rho\equiv w_+/w_-$, 
one sees that for 
 $0\leq\tp\leq1$, $\Te/T$ can acquire any value $\geq
1/2$. If we allow the somewhat more extreme structural choice of $\tp>1$
(thus assuming that both forward and backward steps are promoted with
increasing force) then even negative values of
the effective temperature become possible.
Clearly, even this simple example demonstrates that the idea of 
phenomenologically characterizing
active processes by an elevated ``effective temperature'' is not
really consistent. It rather conceals the physically transparent
additive relationship between mobility and dispersion
by replacing it with a multiplicative factor.

Rather than looking at the response of the motor to a changing applied
force, one can  ask for the response to a change in concentration of
ATP or ADP, i.e., to a change in the chemical potential with
$h_\beta\equiv \mu_{\beta}$.
For the current, we can either choose the ordinary spatial
 current $j(t)$  or the current of  consumed 
$\alpha$-molecules $j_\alpha(t)$. How the corresponding
mean currents change with the 
chemical  potential of $\beta$-molecules
is expressed by the mobility tensor
$
\kappa_{\alpha\beta}$ shown in Table I which includes the ``cross''
mobilities between chemical and mechanical (here denote by an index $f$)
distances and fields.
 We refrain from listing the dispersions, which are in this case symmetrical
with $D_{\alpha\beta}=D_{\beta\alpha}$, and the
corresponding effective temperatures $T^{\rm eff}_{\alpha\beta}$
except for pointing out that the latter are asymmetric and
depend on the choice of indices even for fixed rates.

\begin{table}
\begin{tabular}{|c|c|c|c|c|c|c|c}
\hline
$T \kappa_{\alpha\beta}$ & $\beta $ & $f$ & $t$ & $d$ \\
\hline
$\alpha$& &&&\\
\hline
$f$ & &$d^2(\tp w_++\tm w_-)$ &  $d w_+$ & $-d w_-$\\
\hline
$t$ &&$d(\tp w_+ + \tm w_-)$ & $w_+$&$-w_-$\\
\hline
$d$&&  $-d(\tp w_+ + \tm w_-)$ & $-w_+$&$w_-$\\
\hline

\end{tabular}

\caption{Generalized mobilities for the one-state motor.} 
\end{table}

While the evaluation of mobilities and dispersions is straightforward
also for any more complex specific model as will be illustrated elsewhere, 
a few universal statements seem to be
possible beyond the obvious ones refering to equilibrium. As one example
 consider the observation made in \cite{lau07a} for a particular
two state motor model that at
stalling conditions, $j=0$ at $f=f_s$, the usual Einstein relation between
mobility and diffusion constant holds true, even though idle chemical 
currents dissipate energy. Our expressions (\ref{eq:nu}),
(\ref{eq:I}) and (\ref{eq:Ilocab}) 
show  that, in general, the 
validity of the Einstein relation requires not only that $j=0$ but
moreover that any link carrying a non-zero probability current $K_{mn}$
be not sensitive to the force $f$, i.e. for any $K_{mn}\not = 0$,
$r_{mn}=r_{nm}=0$
must hold. The latter 
condition will not necessarily be met at stalling since even pure chemical
transitions with $d_{mn}=0$ will, in general,  be affected by 
changing the applied force.

{\sl Proof of (\ref{eq:kappamain}).--} 
In the differential mobility 
\begin{eqnarray}
\kappa_{\alpha\beta} &\equiv&  \partial_{h_\beta} j_\alpha =   
\sum_{mn}   d^\alpha_{mn} \partial_
{ h_\beta}
( p_m\wmn) \\ &=&
 \sum_{mn}   d^\alpha_{mn} (\partial_
{ h_\beta}
 p_m) \wmn  + \kappa_{\alpha\beta}^{\rm loc}   
\label{eq:kappa}
\end{eqnarray}
the term  $\partial_
{ h_\beta}
p_m \equiv \partial_{h_\beta} \langle \delta_{n(t)m}\rangle$ 
must be expressed by a correlation function. In \cite{seif09}  we have determined
 the response
of an observable at time $t_2$
 to a delta-like perturbation at time $t_1$.
Specialized to the present quantities and slightly adapting the notation, 
this relation reads
\beq
\left.\frac{\delta \langle \delta_{n(t_2)m}\rangle}{\delta h_\beta(t_1)}\right|_{\{h_\beta\}=const}
= \langle \delta_{n(t_2)m} B (t_1)\rangle  ,
\label{eq:fdt} 
\ee
where the conjugate variable $B(t_1)$ is given by
\begin{eqnarray}
T B(t_1)\equiv \sum_j\delta(t_1-t_j) r^\beta_{n_j^-n_j^+} -
  \sum_k w_{n(t_1)k}r^\beta_{n(t_1)k}~~ \\
= j_\beta(t_1) + \sum_j\delta(t_1-t_j) r^\beta_{n_j^+n_j^-}    -
  \sum_k w_{n(t_1)k}r^\beta_{n(t_1)k}~~ ,
\end{eqnarray}
where we have used (\ref{eq:j}) and (\ref{eq:sym}).
If the correlation function
$
 \langle \delta_{n(t_2)m}  \sum_j\delta(t_1-t_j) r^\beta_{n_j^+n_j^-}\rangle
$
is averaged over the states $n_j^-$ before the jump at $t_1$ one gets
\begin{eqnarray}
 \langle \delta_{n(t_2)m}\sum_j \delta(t_1-t_j) r^\beta_{n_j^+n_j^-}\rangle
~~~~~~~~~~~~~~~~~~~~~~~~~~~~~~
\\=\langle  \delta_{n(t_2)m} \sum_k p_kw_{kn(t_1)}
r^\beta_{n(t_1)k}/p_{n(t_1)}\rangle   . ~~~~~~~~~
\label{eq:bp}
\end{eqnarray}
Putting together (\ref{eq:fdt}-\ref{eq:bp})  and using  (\ref{eq:nu}),
we can write
\beq
T\left. \frac{\delta \langle \delta_{n(t_2)m}\rangle}{\delta h_\beta(t_1)}\right|_{\{h_\beta\}=const}
= \langle \delta_{n(t_2)m} (j_\beta (t_1) -\nu_\beta(t_1))\rangle .
\ee
Thus the response of the current at the later time, 
$
j_\alpha(t_2) = \sum_{nl}
\delta_{n(t_2)n} d^\alpha_{nl}w_{nl}$, 
 to a
delta-like perturbation at the earlier time can be expressed as
\beq
T \left.\frac{\delta \langle j_\alpha (t_2)\rangle}{\delta h_\beta(t_1)}\right|_{\{h_\beta\}=const}
= \langle j_\alpha(t_2) (j_\beta(t_1)-\nu_\beta(t_1)\rangle  .
\ee 
Integrating over the time-difference $t_2-t_1$,
 we obtain  our main result (\ref{eq:kappamain}).

{\sl Concluding perspective.--}
We have expressed the generalized  mobilities by current
correlation
functions for any driven
system described by a master equation with transition rates which obey 
a local detailed balance condition as it should hold for transport in a
well-defined
thermal environment. Without this condition one could still express
the mobility by an integral over some correlation function as a minor
modification of our proof would show. 
The physically transparent connection to the dispersions emphasized here, 
however, would then be lost.  Even though our relation is
remarkably reminiscent to the
well-known linear response result, a crucial difference
should not go unnoticed. For a non-equilibrium steady state as investigated
here, the relevant correlations involve  a ``sensitivity-weighted'' current.
As an observable, the latter requires knowledge of how the rates
depend on the external perturbation. While this is not an issue in any
theoretical modelling, it will limit the direct
application to
those experimental systems for which this property of the rates is
 accessible. In the familiar linear response realm
of the regular Green-Kubo relations, such explicite knowledge is not 
necessary. This observation might support the view that often 
the quantitative  evaluation of exact
non-equilibrium relations requires more specific
input  than their equilibrium counter-parts do.

{\sl Acknowledgments.--} 
An inspiring collaboration with T. Speck on preceeding projects and
financial support by the DFG (through SE1119/3) and ESF (through EPSD) 
are gratefully acknowledged.

\end{document}